\documentclass[11pt,letterpaper]{article}
\usepackage{systeme} %para usar sistemas de ecuaciones
\usepackage[utf8]{inputenc}
\usepackage{multirow} %para las tablas
\usepackage{booktabs} %para la estética de tablas
\usepackage[english]{babel}
\usepackage{amsmath}
\usepackage{amsfonts}
\usepackage{amssymb}
\usepackage{graphicx}
\usepackage[small,bf]{caption} %Para caption con letra pequeña y Figura en negrita
\usepackage[left=3cm,right=3cm,top=2cm,bottom=3cm]{geometry}
\usepackage{framed} %Para cuadro de texto
\usepackage{color} %Para cuadro de texto
\usepackage{wrapfig}\definecolor{shadecolor}{RGB}{220,220,220} %Para color cuadro de texto, codigo 220,220,220 es gris

\author{Jorge Pinochet}
\title{\textbf{Illuminating dark matter I: A guide for physics teachers}}
\begin{document}

\author{Jorge Pinochet$^{*}$\\ \\
 \small{$^{*}$\textit{Facultad de Ciencias Básicas, Departamento de Física. }}\\
  \small{\textit{Centro de Investigación en Educación (CIE-UMCE),}}\\
 \small{\textit{Núcleo Pensamiento Computacional y Educación para el Desarrollo Sostenible (NuCES).}}\\
 \small{\textit{Universidad Metropolitana de Ciencias de la Educación,}}\\
 \small{\textit{Av. José Pedro Alessandri 774, Ñuñoa, Santiago, Chile.}}\\
 \small{e-mail: jorge.pinochet@umce.cl}\\}

\date{}
\maketitle

\begin{center}\rule{0.9\textwidth}{0.1mm} \end{center}
\begin{abstract}
\noindent One of the great mysteries of contemporary science is dark matter, an exotic substance of unknown nature that, in theory, makes up about 27\% of the total mass-energy density of the universe, and which does not appear to emit, absorb, or reflect any kind of light, meaning that it is invisible and can only be detected through its gravitational effects on objects around it. Dark matter is a frontier topic, involving highly complex subjects that usually exceed the training of a physics teacher. Given this difficulty, the aim of this paper is to shed some light on dark matter, and to offer a broad, up-to-date introduction that is mainly directed at physics teachers in training and in practice. Due to the breadth of the subject, the article has been divided into two parts. In Part I, we deal with general concepts, which serve as an introduction to the more specific topics analysed in Part II.\\ \\


\begin{center}\rule{0.9\textwidth}{0.1mm} \end{center}
\end{abstract}

\maketitle

\section{Introduction}

One of the most fascinating mysteries in contemporary physics and astronomy involves dark matter, an exotic substance of unknown nature that is believed to be present throughout the universe [1]. Dark matter makes up about 27\% of the total observed mass-energy density, and does not appear to emit, absorb, or reflect any kind of light, which means that it is invisible and therefore can only be detected indirectly through the gravitational attraction that it produces on the objects in its surroundings [2]. The other 73\% of the mass-energy density is roughly divided into 5\% ordinary matter and 68\% dark energy. The latter is another mysterious substance that should not be confused with dark matter; instead of gravitationally attracting the objects that are in its surroundings, it generates a repulsive gravitational effect that is responsible for the accelerated expansion of the universe.\\

Dark matter occupies a central role in our description of the universe on a large scale, and is a fundamental ingredient of the Standard Cosmological Model, also known as the \textit{Lambda-Cold Dark Matter Model} ($\Lambda$CDM)\footnote{The Greek letter $\Lambda$ (lambda) symbolizes a component of the model called \textit{cosmological constant}, which manifests itself as a form of energy that uniformly fills the entire universe. Due to the unknown nature of this energy, it is also called \textit{dark energy}. As we noted at the end of the first paragraph, it is important not to confuse dark energy with dark matter. The cosmological constant was originally introduced by Einstein, and explains the accelerated expansion of the universe, discovered through astronomical observations in 1998.} [3]. Hence, one of the most pressing problems in contemporary physics is to unravel the true nature of dark matter. Despite the theoretical and experimental efforts that have been made for decades, and the important financial resources invested in this search, to date no one has been able to unveil the mystery of dark matter.\\

Dark matter is a frontier topic, meaning that it involves highly complex subjects that usually exceed the training of a physics teachers. These topics include the theory of general relativity, cosmology, and particle physics, to name only a few. As a result, many high school physics and astronomy teachers are dealing with dark matter on basically the same footing as the general public. In view of this difficulty, the aim of this paper is to shed some light on dark matter, offering a broad and up-to-date introduction that is mainly directed towards physics teachers in training and in practice. To achieve this goal, we will omit all the complexities and technicalities as far as possible, but without renouncing mathematical calculations or scientific rigor. As the subject is very extensive, we will need to be selective. Hence, we do not intend to present a detailed analysis of the problem of dark matter, but seek only to provide material that can serve as a guide for physics teachers.\\

Due to the breadth of the subject, the article has been divided into two parts. In Part I, we introduce some general concepts that will serve as an introduction to the more specific topics analyzed in Part II. Section 1 presents a brief conceptual and historical synopsis of dark matter. In Section 2, we develop a simple Newtonian model that allows us to reproduce the astronomical observations that suggest the existence of dark matter in galaxies. Finally, we will analyze what it means, from an astronomical point of view, that dark matter represents 27\% of the mass-energy density of the universe.

\section{Dark matter: Brief conceptual and historical synthesis}

The first researcher to suggest the presence of invisible matter between galaxies, or \textit{dark matter} as it is known today, was the astrophysicist Fritz Zwicky in 1933 [4–6]. However, Zwicky was unable to convince his colleagues of the importance of his discovery, which remained forgotten for almost forty years. In the 1970s, astronomical observations by Vera Rubin and Kent Ford provided the first strong evidence for the existence of dark matter from an analysis of the \textit{rotation curves} of neighbouring spiral galaxies [4,7]. A rotation curve is a graph of the speed of rotation of visible matter in a galaxy as a function of its radial distance from the centre.\\

A spiral galaxy consists of a flat disc made up of gas, dust, and young stars, a central region or \textit{bulge} made up of a large concentration of old stars that typically revolve around a supermassive black hole, and a spheroidal region made up of old stars called the  \textit{halo}, which surrounds the disc and the bulge [8]. Using this information, and in order to avoid unnecessarily complicating the analysis, we can imagine that the galaxies studied by Rubin and Ford are spheres of mass $M$, radius $R$, and uniform density $\rho$, as illustrated in Fig. 1. As we will see in detail in the next section, this simple model contains all the fundamental aspects of the problem at hand, and allows the findings of Rubin and Ford to be easily explained. Let's analyse Fig. 2, which shows a rotation curve for our spherical galaxy, that is, a graph that describes the rotation speeds of the stars around the centre of the spherical mass distribution.\\

\begin{figure}[h]
  \centering
    \includegraphics[width=0.25\textwidth]{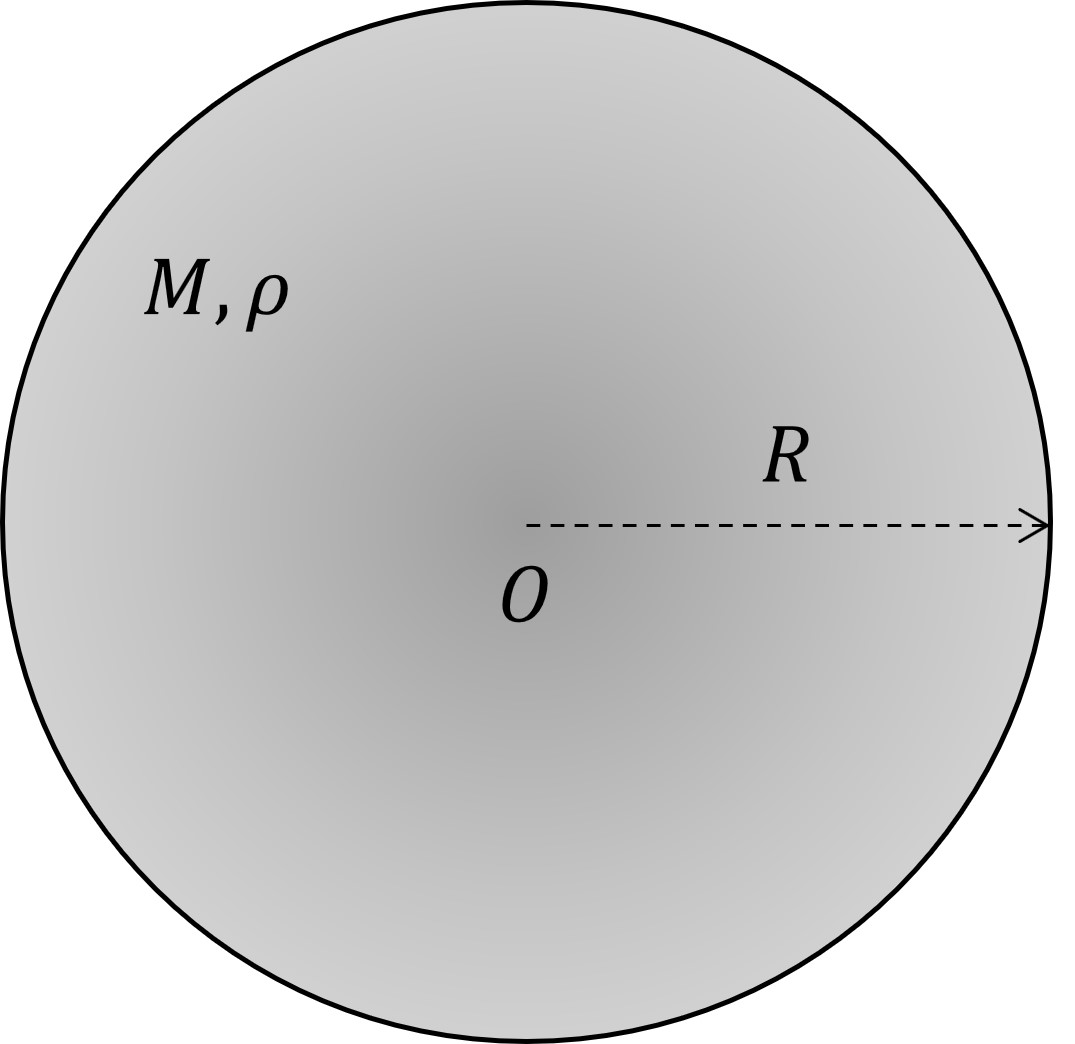}
  \caption{A galaxy modelled as a sphere of radius $R$, mass $M$, and uniform density $\rho$ of visible matter.}
\end{figure}

Suppose that almost all the mass of the galaxy is contained in a sphere of mass $M$ and radius $R$, in an analogous way to our solar system, where 99.9\% of the mass is made up of a spherical central object, the Sun. As we will discuss in more detail in the following sections, applying Newton's law of gravitation and Kepler's laws in the region $r\leq R$ (inside the galaxy) should yield the sloping red line in Fig. 2, which shows that the speed of rotation of stars and visible matter increases linearly with radial distance. Applying the same laws to the region $r>R$ (outside the galaxy), where there is a distribution of neutral hydrogen gas whose rotation speed can be determined (for example, by spectrographs), should give the blue dashed curve, called the \textit{Keplerian rotation curve}, where the speed decreases with distance as $r^{1/2}$, analogous to what happens to the speed of rotation of the planets around the Sun. This is what Rubin and Ford hoped to find by analysing the rotation curves of neighboring galaxies. Although what they found for the region $r\leq R$ matched expectations based on Newtonian theory, what they found for $r>R$ was a \textit{flat rotation curve}, represented in Fig. 2 by a horizontal red line, where the rotation speed was constant over a large range of values beyond the distribution of most visible matter, regardless of the distance from the centre $O$.\\

\begin{figure}[h]
  \centering
    \includegraphics[width=0.5\textwidth]{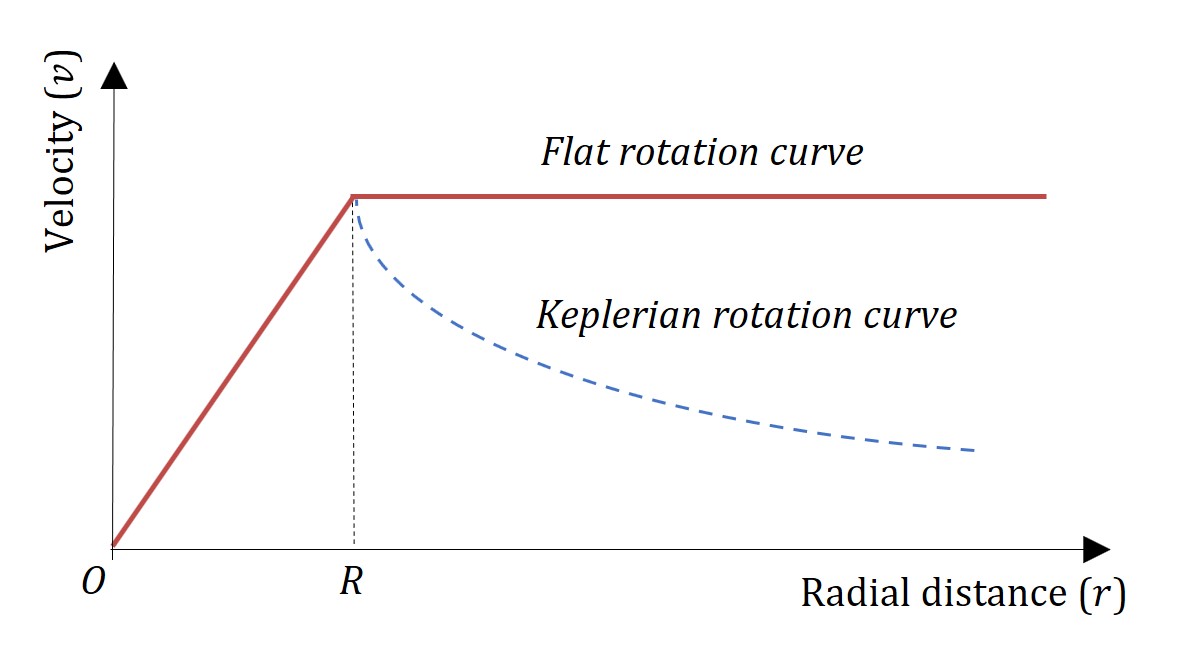}
  \caption{Flat and Keplerian rotation curves for a spherical galaxy of uniform density.}
\end{figure}

For the reader who is curious about what a more realistic rotation curve looks like, i.e., closer to the results obtained by Rubin and Ford, Fig. 3 shows a typical curve. It can be seen that in neither of the two regions is the curve strictly straight. In particular, in the region $r>R$, the curve is not completely flat as shown in Fig. 2, but is approximately flat. In any case, it is important to note that our simple spherical galaxy model reproduces all the fundamental aspects of the findings of Rubin and Ford, and the key point is that according to Figs. 2 and 3, the speed described by the flat curve for $r>R$ is greater than that predicted by the Keplerian curve. According to classical Newtonian dynamics, and in particular according to Newton's law of gravitation, a higher rotation speed implies a greater gravitational mass, which suggests that there is a component of non-visible matter in the region $r>R$. In other words, the visible matter of the galaxy represents only a fraction of the total mass, and the additional mass that is not visible is distributed mainly in the region $r>R$. This was found by Rubin and Ford when they analyzed the rotation curves of the neighbouring galaxies, which represented a revolutionary discovery. Subsequent research using other observational approaches has confirmed this great finding [4].\\

\begin{figure}[h]
  \centering
    \includegraphics[width=0.5\textwidth]{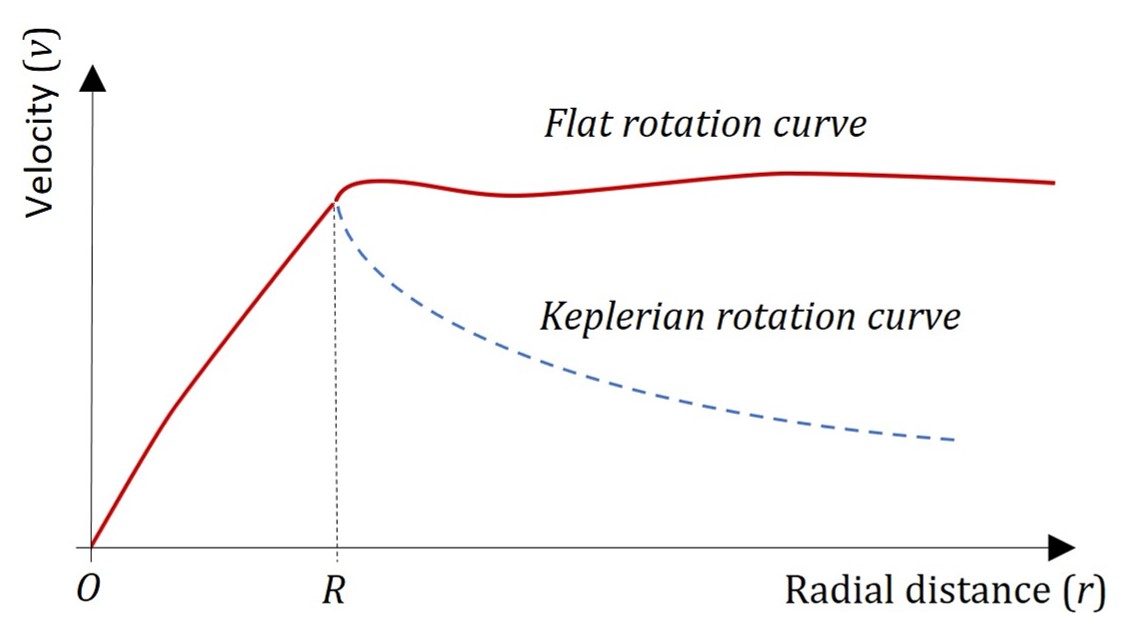}
  \caption{A typical rotation curve obtained from observations of spiral galaxies.}
\end{figure}

Fortunately, one of the approaches that has provided some of the strongest evidence for the existence of dark matter is easy to explain and understand. This is the \textit{gravitational lensing effect}, a prediction of Einstein's theory of gravity, the general relativity. Gravitational lensing is the deflection of light rays from a distant source when they pass near a gravitating mass distribution that acts as a lens [8,9]. Hunters of dark matter have used this effect by studying the bending of light from distant background (source) galaxies as it passes close to a galaxy cluster (lens), as illustrated in Fig. 4, which shows a simple situation with just a background galaxy generating two images. Gravitational lensing can generate multiple images of the source (galaxies), which can take the form of elongated, curved arcs. If the source, lens, and observer are perfectly aligned, a complete luminous ring known as an \textit{Einstein ring} can appear.\\

\begin{figure}[h]
  \centering
    \includegraphics[width=0.6\textwidth]{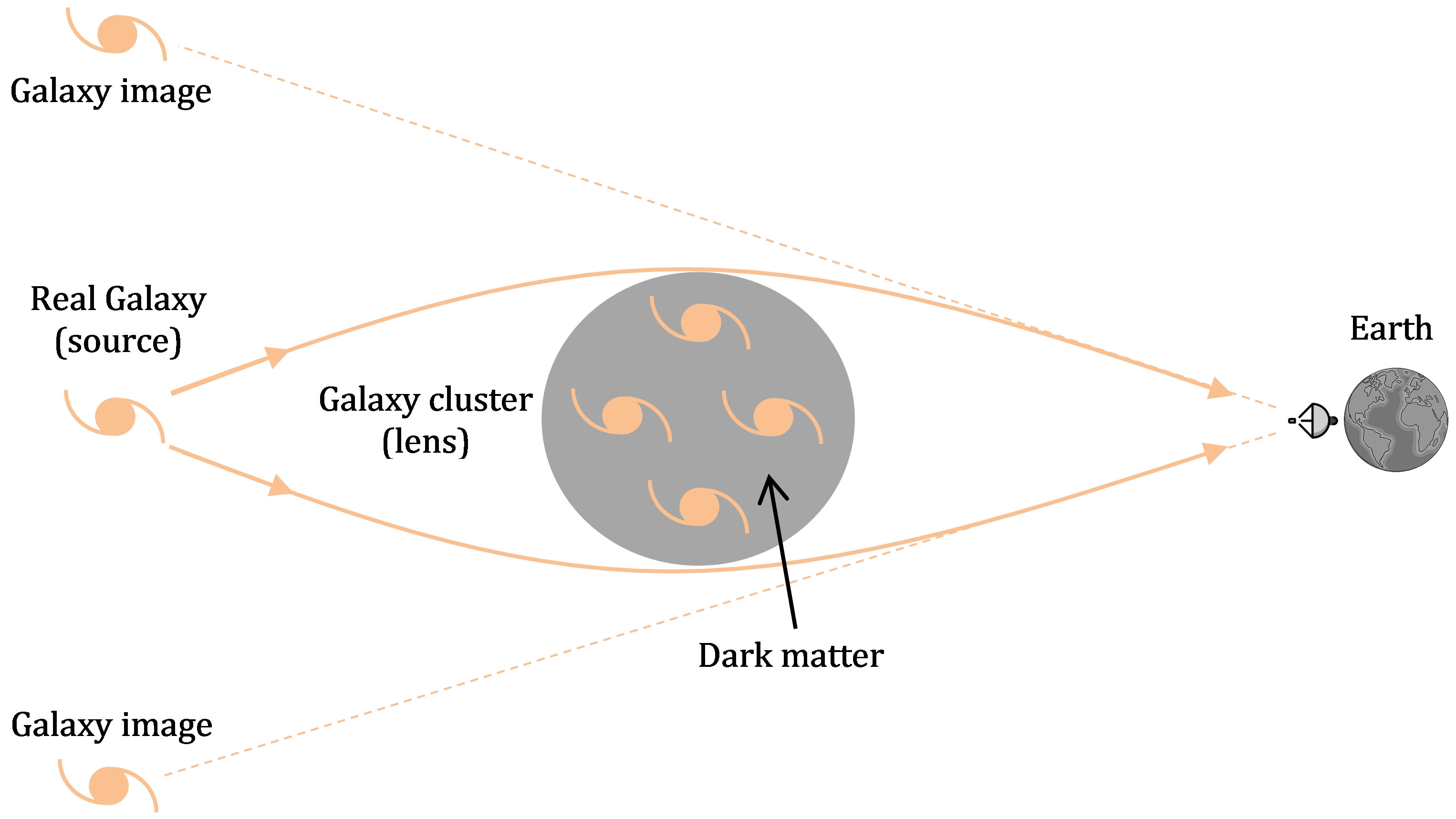}
  \caption{Light from distant background galaxies (source) is deflected as it passes near a cluster of galaxies (lens). If there is dark matter in the cluster, the observed deviation will be greater than that expected from the visible matter. This representation is a particular case when source-lens-observer are aligned, and in this case an Einstein ring is observed.}
\end{figure}

Fig. 5 shows the gravitational lensing effect caused by the galaxy cluster Abell 370, located about 4 billion light-years away. In between the galaxies are arcs of blue light, which are distorted images of remote galaxies behind the cluster, formed by gravitational lensing. By examining these images, and applying the equations of general relativity, it is possible to detect the presence of dark matter and determine its spatial distribution and mass [3]. Similarly to the rotation curves method, analysis of gravitational lensing has revealed that the visible mass of galaxies and clusters of galaxies represents only a fraction of the total mass.\\

Current observations suggest that dark matter represents approximately 84\% of the matter in the universe and 27\% of its total mass-energy density [2]. Dark matter does not show any interaction with itself or with electromagnetic radiation (such as light), and is therefore very difficult to detect. Initially, the simplest and most conservative solution was chosen, which was the assumption that dark matter was composed of \textit{baryons} (the family of subatomic particles to which protons and neutrons belong), and was therefore ordinary matter that makes up everything around us. However, current observations favour models where the primary component is \textit{non-baryonic matter}, that is, exotic matter that is not composed of baryons. These observations also favour cold dark matter (CDM) models. As an analogy with statistical thermodynamics, where the temperature of a gas is proportional to the square of the average velocity (mean kinetic energy) of its constituent particles, the term "cold" is used to indicate that the characteristic velocity of the dark matter particles is small compared to the speed of light in a vacuum, $c=3 \times 10^{8} m/s$. In contrast, hot dark matter (HDM) would be composed of particles travelling at relativistic speeds (close to $c$).\\

\begin{figure}[h]
  \centering
    \includegraphics[width=0.55\textwidth]{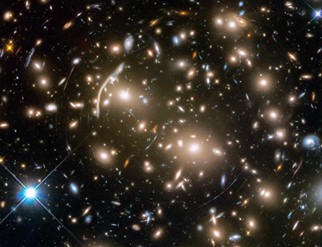}
  \caption{The galaxy cluster Abell 370, located about 4 billion light-years away, causes a gravitational lensing effect on background galaxies, forming distorted images that take the form of arcs of blue light (from: https:// hubblesite.org/).}
\end{figure}

Dark matter plays a fundamental role in the formation of structures such as galaxies or clusters of galaxies\footnote{A cluster of galaxies is a structure composed of hundreds or thousands of galaxies held together by their own gravity, with typical masses ranging between $10^{14}$ and $10^{15}$ solar masses.}. Since it represents about 84\% of the matter in the universe, its strong gravity is crucial in allowing matter to aggregate in the ways that astronomical observations reveal to us. This explains why the empirical evidence favours cold dark matter models. If the primary components were hot dark matter particles, they could easily exceed the escape speed of aggregates of matter in formation, thus preventing the appearance of galaxies and clusters of galaxies. We can see that if dark matter were hot, the universe would be very different.

\section{A simple Newtonian model for dark matter}

To explain the findings of Rubin and Ford in quantitative terms, we return to the spherical galaxy model from the previous section. Recall that according to Newton's law of universal gravitation, the magnitude of the force $F$ exerted by a spherically symmetric distribution of mass $M$ on a particle of mass $m$ is

\begin{equation} %eq1
F = \frac{GMm}{r^{2}},
\end{equation}

where $r$ is the distance between the particle and the centre of the sphere. Let us consider a star of mass $m\ll M$, located inside the spherical galaxy, at an arbitrary distance $r<R$ from its centre (see Fig. 6). According to Gauss's law for gravity, the gravitational force on the star is only due to the mass contained in the region of radius $r$ [10], and the force exerted by the mass contained in the outer region is zero. This means that we can calculate the force of gravity on the star using Eq. (1), considering only the mass $M$ contained in the region of radius $r$; however, we must bear in mind that $M=M(r)$, that is, $M$ increases with $r$, and outside the spherical galaxy (where $r>R$), $M$ is constant. To find the dependence of the mass on the radius, we can express the mass as the product of the density $\rho$, which we assume to be uniform, and the corresponding volume $V$, that is, $M(r)=\rho V = 4\pi r^{3} \rho /3$, such that:

\begin{equation} %eq2
F = \frac{GM(r)m}{r^{2}} = \left(  \frac{4\pi Gm\rho}{3} \right) r.
\end{equation}

Thus, inside the spherical galaxy, gravity varies linearly with the radial distance $r$. For our sphere to resemble a real galaxy, we must assume that the matter contained in it, both baryonic and non-baryonic, describes circular orbits around the centre. To describe this mathematically, we analyse Fig. 6. Due to the circular orbits, a star of mass $m$ located inside the sphere experiences a centripetal force of magnitude

\begin{equation} %eq3
F_{c} = \frac{mv^{2}}{r}.
\end{equation}

This force is due to gravity exerted by the mass contained in the region of radius $r$; then, from Eqs. (2) and (3), we have $F = F_{c}$, which gives

\begin{equation} %eq4
v(r) = \left(  \frac{4\pi G\rho}{3} \right)^{1/2} r = (constant) r.
\end{equation}

As in Eq. (2), the speed of rotation increases linearly with the radial distance. Outside of the spherical galaxy ($r>R$), the force of gravity on the star of mass $m$ is also given by Eq. (1), except that $M$ is constant. Taking $F = F_{c}$ in Eqs. (1) and (3) gives

\begin{equation} %eq5
v(r) = \left(  \frac{GM}{r} \right)^{1/2} = \dfrac{(constant)}{r^{1/2}}.
\end{equation}

\begin{figure}[h]
  \centering
    \includegraphics[width=0.25\textwidth]{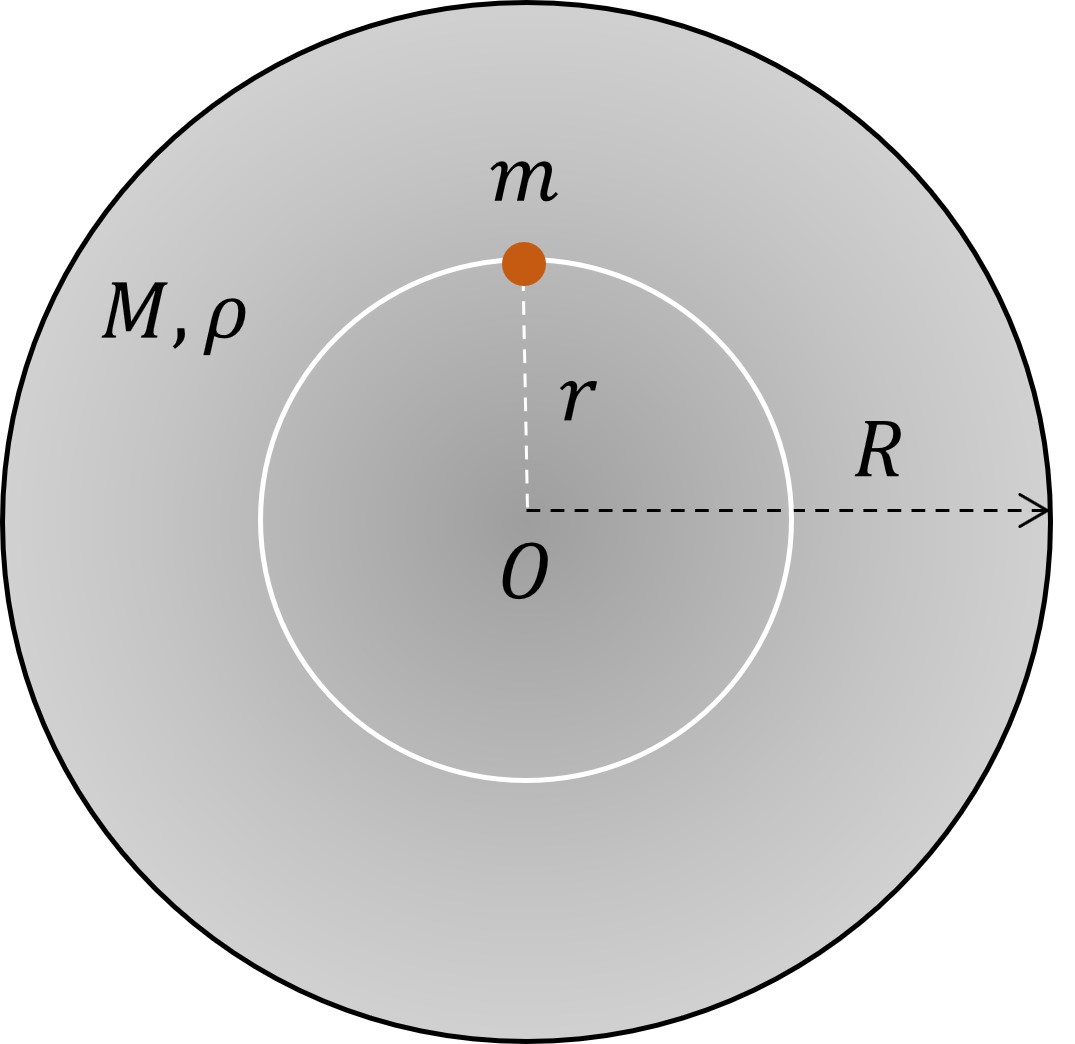}
  \caption{Inside the spherical galaxy, at a distance $r<R$, is a star of mass $m\ll M$.}
\end{figure}

From Eqs. (4) and (5), we can describe the speed of rotation as a piecewise function:

\begin{equation} %eq6
v(r) =
\begin{cases}
\left( \frac{4\pi G \rho}{3} \right)^{1/2} r & \text{if $r \leq R$ (\textit{within the galaxy})} \\

\left( \frac{GM}{r} \right)^{1/2} & \text{if $r > R$ (\textit{out of the galaxy})}.
\end{cases}
\end{equation}

A graph of Eq. (6) gives the Keplerian rotation curve that appears to the left of Fig. 7. The shape of this graph coincides with the Keplerian curve in Fig. 2. To obtain the corresponding flat rotation curve, we must assume that for $r>R$, the speed does not decrease as $1/r^{1/2}$ but is constant. In the framework of classical Newtonian dynamics, this is only possible if in Eq. (5) we take $M(r)=kr$, where $k$ is a constant. This means that in the region $r>R$ there is a non-visible mass distribution that increases linearly with radial distance, which implies that the speed for $r>R$ is constant:

\begin{equation} %eq7
v(r) = \left( \frac{GM(r)}{r} \right)^{1/2} = \left( \frac{Gkr}{r} \right) = \left( Gk \right)^{1/2} = constant.  
\end{equation}

Thus, the speed of rotation is written as

\begin{equation} %eq8
v(r) =
\begin{cases}
\left( \frac{4\pi G \rho}{3} \right)^{1/2} r & \text{if $r \leq R$ (\textit{within the galaxy})} \\

\left( Gk \right)^{1/2} = constant  & \text{if $r > R$ (\textit{out of the galaxy})}.
\end{cases}
\end{equation}

The graph on the right of Fig. 7 shows the curve associated with this expression, which coincides with the flat rotation curve in Fig. 2. We see then that by using classical Newtonian dynamics, we can reproduce the flat rotation curve only if we assume the presence of non-visible matter or dark matter.

\begin{figure}[h]
  \centering
    \includegraphics[width=0.6\textwidth]{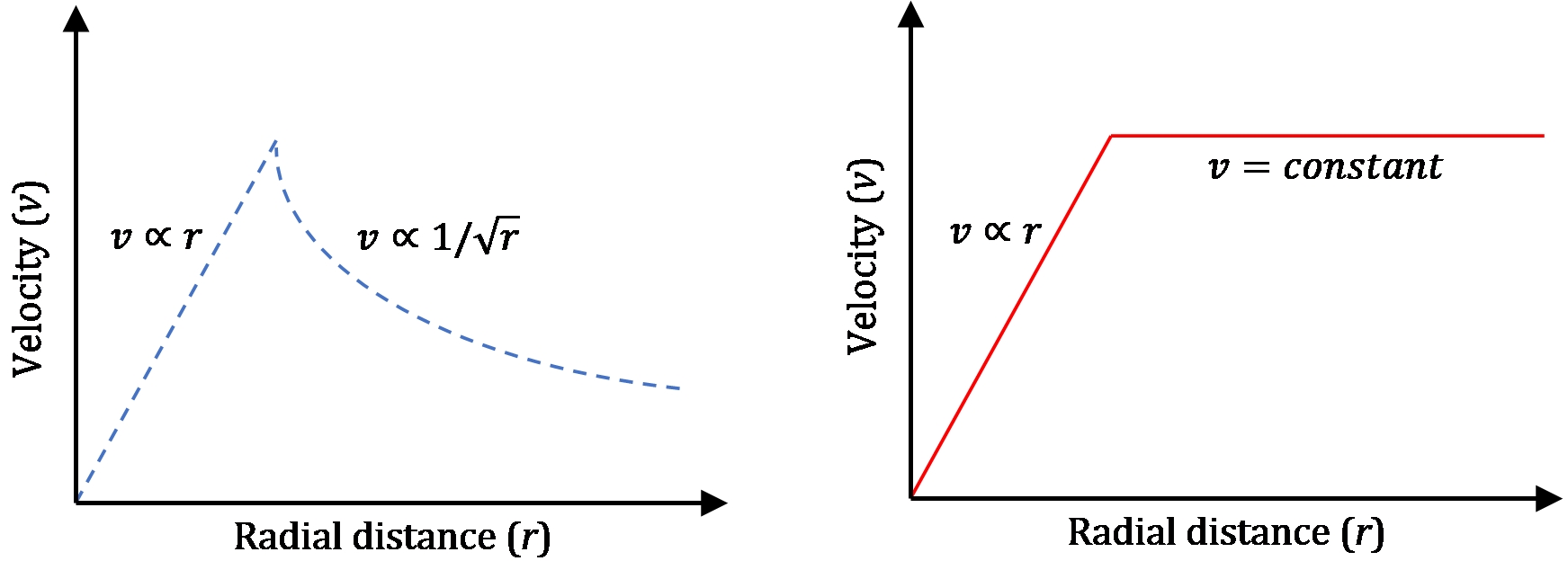}
  \caption{Left: Keplerian rotation curve obtained from the spherical galaxy model. Right: Corresponding flat rotation curve.}
\end{figure}

\section{Dark matter and the density parameter}

Looking in all directions, the galaxies are receding from us, as evidenced by the redshift of their luminous spectra, a discovery made by astronomers Vesto Slipher and Edwin Hubble in the early 20th century [11,12], and predicted theoretically by cosmologist Georges Lemaître. That is why this phenomenon is known as the \textit{Hubble–Lemaître law}. This law provided the first evidence for the Big Bang and the expansion of the Universe, which led to the development of the simplest cosmological model that is in agreement with astronomical observations, the $\Lambda$CDM model, of which cold dark matter is a fundamental ingredient.\\

\begin{figure}[h]
  \centering
    \includegraphics[width=0.4\textwidth]{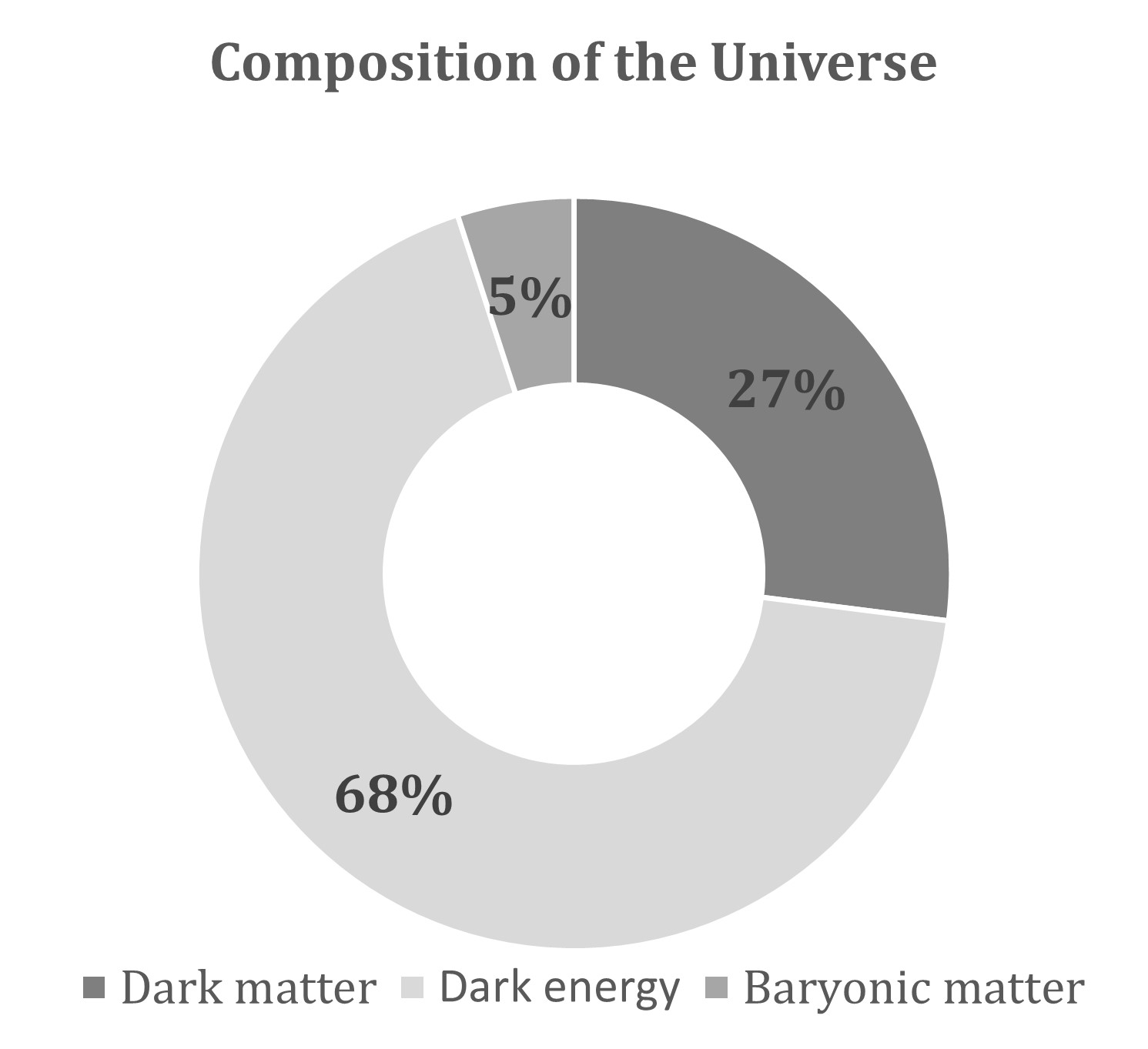}
  \caption{Percentage distribution of the mass-energy associated with the three components of the universe.}
\end{figure}

In cosmology, it is customary to express the mass-energy density of the universe in units of the so-called \textit{critical density}, $\rho_{c} = 9.47\times 10^{-27} kg\cdot m^{-3}$, which is equivalent to 5 atomic nuclei of hydrogen within a cubic meter. From the critical density, we can define a dimensionless quantity called the \textit{total density parameter} [8]:

\begin{equation}%eq9
\Omega_{Tot} \equiv \frac{\rho_{Tot}}{\rho_{c}},
\end{equation}

where $\rho_{Tot}$ is the mass-energy density of the universe, which includes all known forms of matter and energy. According to general relativity, in which the Newtonian attractive forces are replaced by the concept of curvature of space-time, the value of $\rho_{Tot}$ determines the curvature of the universe, and hence establishes the global geometry of space on a large scale. When $\rho_{Tot} =1$, the universe is flat and therefore globally obeys the familiar rules of Euclidean geometry; in contrast, when $\rho_{Tot} \neq 1$, the universe is curved, and its geometry is not governed by Euclid's rules. By definition, $\rho_{c}$ is the value of the mass-energy density for which $\Omega_{Tot} =1$, which implies that $\rho_{c} = \rho_{Tot}$. The parameter $\Omega_{Tot}$ can be divided into three dimensionless components [8]:

\begin{equation}%eq10
\Omega_{Tot} = \Omega_{BM} + \Omega_{DE} + \Omega_{DM},
\end{equation}

where $\Omega_{BM}$ is the \textit{baryonic matter density parameter}, $\Omega_{DE}$ is the \textit{dark energy density parameter}, and $\Omega_{DM}$ is the \textit{dark matter density parameter}, the most important quantity for our purposes. Within observational errors, astronomical measurements have revealed that $\Omega_{BM} \cong 0.05$, $\Omega_{DE} \cong 0.68$ and $\Omega_{BM} \cong 0.27$, so that

\begin{equation}%eq11
\Omega_{Tot} \cong 0.05 + 0.68 + 0.27 = 1.
\end{equation}

The observational evidence therefore establishes that the universe has zero global spatial curvature, which means that its geometry is Euclidean.\\

We are now in a position to clarify the physical meaning of the percentage distribution of the three components of the universe, as illustrated in Fig. 8. Since the sum of the components is equal to one, we can consider that this sum corresponds to 100\% of the composition of the universe. Based on this logic, the density of dark matter represents around 27\% of the total mass-energy density, whereas the baryonic matter density represents 5\%, and the dark energy density 68\%. These results provide clear evidence for the vast amount of dark matter in the Universe, which is about six times more abundant than baryonic matter.

\section{Final comments}

The ideas developed in this first part give us a panoramic view of the problem of dark matter. However, we are far from exhausting everything we could say on this subject. Among other things, we are interested in knowing the types of particles that could make up the primary components of dark matter, as well as the experimental methods used to detect it. We are also interested in exploring the so-called \textit{modified gravity theories}, which have tried to explain these astronomical observations without relying on dark matter. These are the main issues that we will address in Part II of this work.

\section*{Acknowledgments}
I would like to thank to Daniela Balieiro for their valuable comments in the writing of this paper. 

\section*{References}

[1] J. Woithe, M. Kersting, Bend it like dark matter!, Phys. Educ. 56 (2021) 035011.

\vspace{2mm}
 
[2] J.P. Luminet, The Dark Matter Enigma, Inference. 5 (2020). 

\vspace{2mm}

[3] K. Garrett, G. Duda, Dark Matter: A Primer, Advances in Astronomy. 2011 (2011) e968283. 

\vspace{2mm}

[4] G. Bertone, D. Hooper, History of dark matter, Rev. Mod. Phys. 90 (2018) 045002.

\vspace{2mm}
 
[5] F. Zwicky, Republication of: The redshift of extragalactic nebulae, General Relativity and Gravitation. 41 (2009) 207–224.

\vspace{2mm}

[6] F. Zwicky, Die Rotverschiebung von extragalaktischen Nebeln, Helvetica Physica Acta. 6 (1933) 110–127.

\vspace{2mm}

[7] V.C. Rubin, W.K. Ford Jr., Rotation of the Andromeda Nebula from a Spectroscopic Survey of Emission Regions, The Astrophysical Journal. 159 (1970) 379–403.

\vspace{2mm}

[8] D. Maoz, Astrophysics in a Nutshell, Princeton University Press, Princeton, 2016.

\vspace{2mm}

[9] J. Pinochet, Einstein ring: Weighing a star with light, Phys. Educ. 53 (2018) 055003.

\vspace{2mm}

[10] G.B. Arfken, H.J. Weber, Mathematical Methods for Physicists, 6th ed., Elsevier, California, 2005.

\vspace{2mm}

[11] H.S. Kragh, Conceptions of cosmos: From myths to the accelerating universe: A history of cosmology, Oxford University Press, Oxford, 2007.

\vspace{2mm}

[12] Hubble, E, A Relation between Distance and Radial Velocity among Extra-Galactic Nebulae, Proceedings of the National Academy of Sciences of the United States of America.  15 (1929) 168-173.

\end{document}